\pdfoutput=1

\documentclass[%
 reprint,
 showpacs,
 nofootinbib,
 amsmath,amssymb,
 aps,
 pra,
 longbibliography
]{revtex4-1}

\usepackage{graphics}
\usepackage{graphicx}
\usepackage{color}
\usepackage{amsmath}
\usepackage{slashed}
\usepackage{epsfig}
\usepackage{hyperref}

\begin{document}

\title{Natural leptogenesis and neutrino masses with two Higgs doublets}

\author{Jackson D. Clarke, Robert Foot, Raymond R. Volkas}
\affiliation{ARC Centre of Excellence for Particle Physics at the Terascale, \\ 
School of Physics, University of Melbourne, 3010, Australia.}

\date{\today}

\begin{abstract}

The minimal Type I see-saw model cannot explain the observed neutrino masses and the
baryon asymmetry of the Universe via hierarchical thermal leptogenesis without ceding naturalness.
We show that this conclusion can be avoided by adding a second Higgs doublet with $\tan\beta\gtrsim 4$.
The models considered naturally accommodate a SM-like Higgs boson, and predict TeV-scale scalar states and
low- to intermediate-scale hierarchical leptogenesis with $10^3\text{ GeV}\lesssim M_{N_1}\lesssim 10^8\text{ GeV}$.

\end{abstract}


\maketitle

\section{Introduction}

The discovery of a neutral Higgs boson at the Large Hadron Collider (LHC) \cite{Aad2012tfa,Chatrchyan2012ufa}
has strengthened the case for the standard model (SM) paradigm of spontaneous electroweak symmetry breaking:
a scalar doublet $\Phi$ gains a vacuum expectation value (vev) 
$\langle\Phi\rangle=v/\sqrt{2}\approx 174$~GeV
by virtue of the potential
\begin{align}
 V_{\text{SM}} = \mu^2\Phi^\dagger\Phi + \lambda \left(\Phi^\dagger\Phi\right)^2 , \label{EqVSM}
\end{align}
with $\mu^2<0$.
As well, the measurement of neutrino oscillations 
\cite{Cleveland1998nv,Fukuda1998mi,Hampel1998xg,Ahmad2001an,Ahmad2002jz,Eguchi2002dm}
suggests that the SM should be extended to incorporate neutrino masses.
A straightforward way to achieve this is to include three right-handed neutrinos.
Then gauge invariance allows two extra renormalisable terms
to be added to the Yukawa Lagrangian,
\begin{align}
 -\Delta\mathcal{L}_Y = (y_\nu)_{ij} \overline{l^i_L}\tilde\Phi\nu^j_R 
 +\frac12 M_i\overline{(\nu^i_R)^c}\nu^i_R +h.c., \label{EqSeesawYuk}
\end{align}
where $l_L=(\nu_L,e_L)^T$, $\tilde\Phi=i\tau_2\Phi^*$,
and $M_i$ are the right-handed neutrino masses.
The SM extended in this way is what we refer to 
as the minimal Type I see-saw model \cite{Minkowski1977sc,Mohapatra1979ia,Yanagida1979as,GellMann1980vs}.

If $y_\nu v \ll M_i$ then the minimal Type I see-saw provides 
an elegant explanation for the smallness of the neutrino masses.
After electroweak symmetry breaking, the neutrino mass matrix is given by the see-saw formula
\begin{align}
 m_\nu = \frac{v^2}{2} y_\nu \mathcal{D}^{-1}_M y_\nu^T, \label{EqSeesaw}
\end{align}
where $\mathcal{D}_M\equiv \text{diag}(M_1,M_2,M_3)$,
suppressed by the presumably large right-handed neutrino mass scale.

Minimal Type I see-saw also provides a mechanism to reproduce the baryon asymmetry of the Universe (BAU).
Fukugita--Yanagida hierarchical thermal leptogenesis \cite{Fukugita1986hr} proceeds via the
CP-violating out-of-equilibrium decays of the lightest right-handed neutrino $N_1$,
creating a lepton asymmetry which is reprocessed into the baryon
sector by the electroweak sphalerons.
Successful hierarchical thermal leptogenesis is possible when the
Davidson--Ibarra bound 
(ensuring enough CP asymmetry in the decays)
is satisfied \cite{Davidson2002qv,Giudice2003jh},
\begin{align}
 M_{N_1} \gtrsim 5\times 10^8\text{ GeV} \left(\frac{v}{246\text{ GeV}}\right)^2, \label{EqDavIbarra}
\end{align}
where $v$ is the vev that enters the see-saw of Eq.~\ref{EqSeesaw}.

\begin{figure}[t]
 \centering
 \includegraphics[width=0.9\columnwidth]{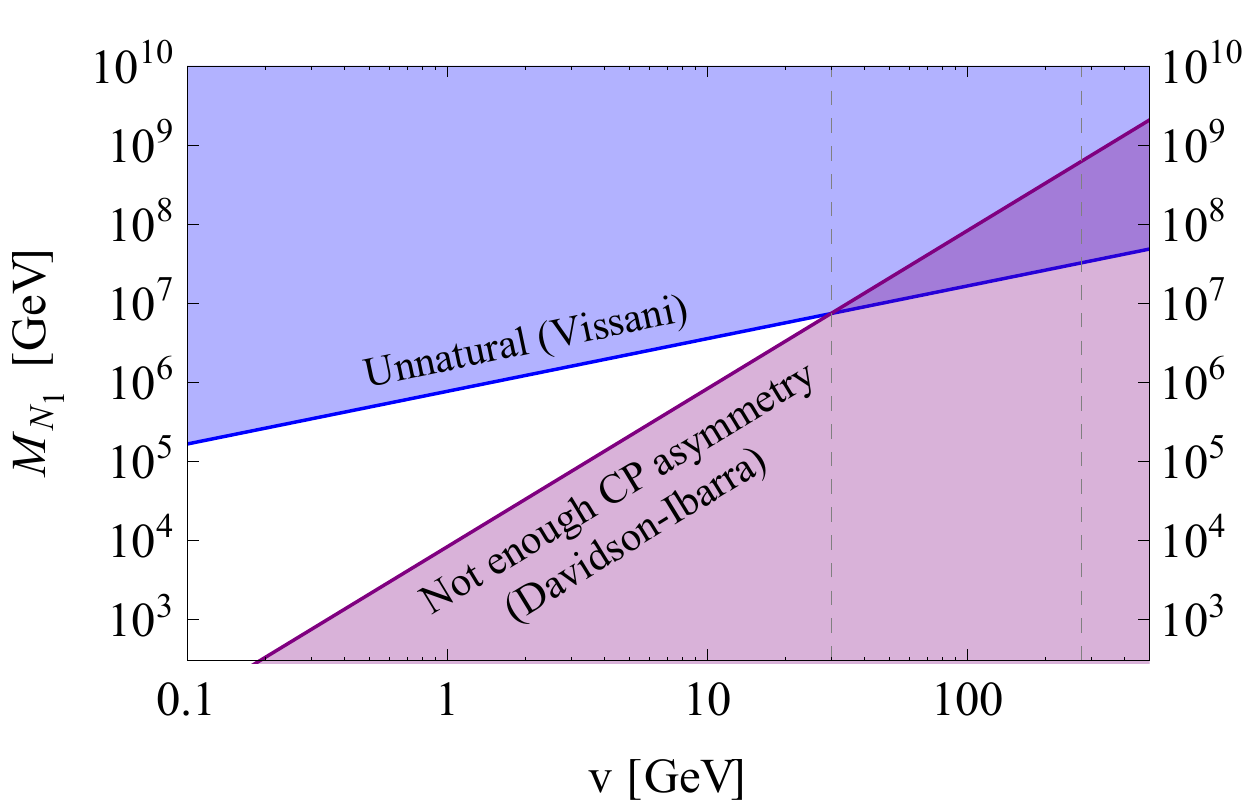}
 \caption{Bounds on hierarchical thermal leptogenesis as a function of $v$.
 Shown is the Davidson-Ibarra bound (purple), and the Vissani bound (blue).
 The dashed lines indicate $v=246,30$~GeV.}
 \label{FigMNv}
\end{figure}

The ability of the minimal Type I see-saw model to simultaneously explain
neutrino masses and the BAU is certainly intriguing.
However, Vissani observed \cite{Vissani1997ys} that the model is incapable of doing so
without generating a naturalness problem.\footnote{Naturalness 
is admittedly an aesthetic requirement of a model,
and the possibility remains that nature is just fine-tuned.} 
Equation~\ref{EqDavIbarra} is simply
incompatible with the conservative naturalness requirement that
corrections to the electroweak $\mu^2$ parameter of Eq.~\ref{EqVSM} 
not exceed 1 TeV$^2$.
With three flavours of hierarchical right-handed neutrinos this requires \cite{Clarke2015gwa}
\begin{align}
 M_{N_1}\lesssim 3\times10^7\text{ GeV}\left(\frac{v}{246\text{ GeV}}\right)^\frac23 . \label{EqNatural}
\end{align}
The incompatibility is exemplified in Fig.~\ref{FigMNv};
nowhere at $v=246$~GeV is it possible to simultaneously fulfil 
the Davidson--Ibarra and Vissani bounds.

A sensible question is then: in what minimal ways can this incompatibility be overcome?
Figure~\ref{FigMNv} suggests three conspicuous (but not mutually exclusive) options:
(1) modify the correction to $\mu^2$, e.g.
by restoring supersymmetry,
or by partly cancelling the correction from
the heavy fermion loop \cite{Bazzocchi2012de,Fabbrichesi2015zna};
(2) lower the Davidson--Ibarra bound, e.g.
by considering resonant leptogenesis \cite{Pilaftsis2003gt},
an alternative mechanism \cite{Akhmedov1998qx},
or by introducing new fields which allow an increased CP asymmetry in the right-handed neutrino decay;
(3) seek an extension of the canonical see-saw for neutrino mass, i.e. 
reduce the (possibly effective) $v$ entering the see-saw Eq.~\ref{EqSeesaw}.

In this paper we will consider the third option.
Specifically we will examine alternative see-saw possibilities when the 
minimal Type I see-saw model is extended by a second Higgs doublet $\Phi_2$.
We are motivated by the following observation: 
if the see-saw neutrino mass of Eq.~\ref{EqSeesaw} is evaluated at $v\lesssim 30$~GeV,
then Eqs.~\ref{EqDavIbarra} and \ref{EqNatural} become compatible,
as is clear from Fig.~\ref{FigMNv}.
Thus we expect that two-Higgs-doublet models with right-handed neutrinos ($\nu$2HDMs) 
and $\tan\beta = v_1/v_2\gtrsim 8$,
where $\Phi_2$ is responsible for a tree-level see-saw,
can naturally accommodate leptogenesis and neutrino masses.
In fact, we find that $\tan\beta \gtrsim 4$ is possible,
since the extra scalar states can be naturally TeV-scale
and the Vissani bound can be relaxed.

Upon examining the $\nu$2HDM scenarios which will succeed,
we rediscover the radiative Ma model \cite{Ma2006km} as the only possibility when $v_2=0$.
Otherwise, in models without a significant radiative neutrino mass component,
we require $0.3 \lesssim v_2/\text{GeV}\lesssim 60$.
The potentially small vev is made natural by softly breaking a $U(1)$ or $Z_2$ symmetry,
which also automatically results in one SM-like CP-even state.
One advantage of this scenario is that it can work in all $\nu$2HDM Types,
greatly increasing the opportunity for model building.

The paper is organised as follows.
In Sec.~\ref{Secnu2hdm} we build the $\nu$2HDM models of interest,
describe the scalar states, and briefly review the relevant experimental constraints.
In Sec.~\ref{Secnu2hdmNat} we pay particular attention to naturalness limits on the extra scalars;
we verify that a natural $\nu$2HDM of any Type is still allowed by experiment.
We discuss neutrino masses in Sec.~\ref{Secnumasses}
and leptogenesis in Sec.~\ref{SecLeptogen}. 
The region of parameter space which naturally achieves
hierarchical leptogenesis is identified.
We conclude in Sec.~\ref{secconclusion}.

\section{The $\nu$2HDM model \label{Secnu2hdm}}

\subsection{Lagrangian}

The scalar content of the model contains two doublets $\Phi_{1,2}$ each with hypercharge +1.
For simplicity we consider the softly broken, CP-conserving, $Z_2$-symmetric potential
(see e.g. Ref.~\cite{Branco2011iw})
\begin{align}
 V_{\text{2HDM}} =&\;
m^2_{11}\, \Phi_1^\dagger \Phi_1
+ m^2_{22}\, \Phi_2^\dagger \Phi_2 
-m^2_{12}\, \left(\Phi_1^\dagger \Phi_2 + \Phi_2^\dagger \Phi_1\right) 
\nonumber\\ &
+ \frac{\lambda_1}{2} \left( \Phi_1^\dagger \Phi_1 \right)^2
+ \frac{\lambda_2}{2} \left( \Phi_2^\dagger \Phi_2 \right)^2
\nonumber\\ &
+ \lambda_3\, \left(\Phi_1^\dagger \Phi_1\right) \left(\Phi_2^\dagger \Phi_2\right)
+ \lambda_4\, \left(\Phi_1^\dagger \Phi_2\right) \left(\Phi_2^\dagger \Phi_1\right)
\nonumber\\ &
+ \frac{\lambda_5}{2} \left[
\left( \Phi_1^\dagger\Phi_2 \right)^2
+ \left( \Phi_2^\dagger\Phi_1 \right)^2 \right], \label{EqV2HDM}
\end{align}
where all the parameters are real.
To explain observations, at least one of these doublets must obtain a non-zero vev.
We consider $m_{11}^2<0$ and a CP-conserving vacuum,
\begin{align}
 \langle\Phi_1\rangle_0 = \frac{1}{\sqrt{2}}\left(
 \begin{array}{c}
  0 \\ v_1
 \end{array}
 \right),
 &&
 \langle\Phi_2\rangle_0 = \frac{1}{\sqrt{2}}\left(
 \begin{array}{c}
  0 \\ v_2
 \end{array}
 \right),
\end{align}
where $v_1>0$, $v_2\ge 0$, and $v_1^2+v_2^2=v^2\approx(246\text{ GeV})^2$.

\begin{table}
 \begin{tabular}{c|c|c|c|c}
  Model 	& $u_R^i$ 	& $d_R^i$ 	& $e_R^i$ 	& $\nu_R^i$ \\
  \hline
  Type I	& $\Phi_1$	& $\Phi_1$	& $\Phi_1$	& $\Phi_2$ \\
  Type II	& $\Phi_1$	& $\Phi_2$	& $\Phi_2$	& $\Phi_2$ \\
  Lepton-specific (LS)	& $\Phi_1$	& $\Phi_1$	& $\Phi_2$	& $\Phi_2$ \\
  Flipped	& $\Phi_1$	& $\Phi_2$	& $\Phi_1$	& $\Phi_2$ \\
 \end{tabular}
 \caption{The four models with no tree-level flavour-changing neutral currents 
 and allowing for a GeV-scale vev to provide the see-saw whilst preserving perturbativity of $y_t$.}
 \label{TabTypes}
\end{table}

A general 2HDM will have flavour-changing neutral currents at tree-level.
These can be avoided if right handed fermions of a given type $(u_R^i,d_R^i,e_R^i)$ 
couple to only one of the doublets \cite{Paschos1976ay,Glashow1976nt}.
Although not strictly necessary, we will assume that this is realised,
and adopt the convention that only $\Phi_1$ couples to the $u_R^i$.
In a $\nu$2HDM, if we assume this also applies for the $\nu_R^i$, 
then there are eight possibilities.
As mentioned in the Introduction, the see-saw constraint Eq.~\ref{EqSeesaw} can be made
consistent with naturalness and leptogenesis if the vev contributing
to the see-saw is sufficiently small.
Since we would like our model to remain perturbative, and already $y_t\approx 1$
for $v\approx 246$~GeV,
we anticipate that $\Phi_2$ obtains the small vev and thus we couple it to the $\nu_R^i$.
Remaining are four possible $\nu$2HDMs which we refer to by their conventional Types
as listed in Table~\ref{TabTypes}.\footnote{Type I
$\nu$2HDMs with $v_2\sim$~eV were considered in Refs.~\cite{Ma2000cc,Wang2006jy,Gabriel2006ns,Davidson2009ha}.
We will end up considering $v_2$ of $\mathcal{O}(0.1$--$10)$~GeV.}
The Yukawa Lagrangian is then given by
\begin{align}
 -\mathcal{L}_Y =&\, + y_u\overline{q_L}\tilde\Phi_1 u_R + y_d\overline{q_L}\Phi_I d_R \nonumber \\
   & + y_e \overline{l_L}\Phi_J e_R + y_\nu \overline{l_L}\tilde\Phi_2 \nu_R \nonumber \\
   & + \frac12 M_N \overline{(\nu_R)^c}\nu_R + h.c. , \label{EqYuk}
\end{align}
where $I,J$ depend on the model Type, and family indices are implied.

\subsection{Scalar masses and mixings \label{Secmassmix}}

Consistency with experiments requires the extra scalar states
to have masses at least $\gtrsim 80$~GeV.
In order to construct models with potentially TeV-scale scalars with a 
naturally small $v_2$, we will consider $m_{22}^2>0$ and $m_{12}^2/m_{22}^2\ll 1$ \cite{Ma2000cc}.
This is technically natural, since in the limit $m_{12}^2/m_{22}^2\to 0$
a $U(1)$ or $Z_2$ symmetry is restored if $\lambda_5=0$
or $\lambda_5\ne 0$, respectively.

For $m_{11}^2<0$, the vevs are given by
\begin{align}
 v_1 &\approx \sqrt{\frac{-2m_{11}^2}{\lambda_1}} , \label{Eqvev1} \\
 v_2 &\approx \frac{1}{1+\frac{v_1^2}{2m_{22}^2}\lambda_{345}} \frac{m_{12}^2}{m_{22}^2}v_1 ,\label{Eqvev2}
\end{align}
where $\lambda_{345}=\lambda_3+\lambda_4+\lambda_5$.
These relations become exact when $m_{12}^2=0$ and when $\lambda_2=0$, respectively.
For $m_{22}^2\gg \lambda_{345} v_1^2$, $\tan\beta \equiv v_1/v_2\approx m_{22}^2/m_{12}^2$.

\begin{figure}[t]
 \centering
 \includegraphics[width=0.9\columnwidth]{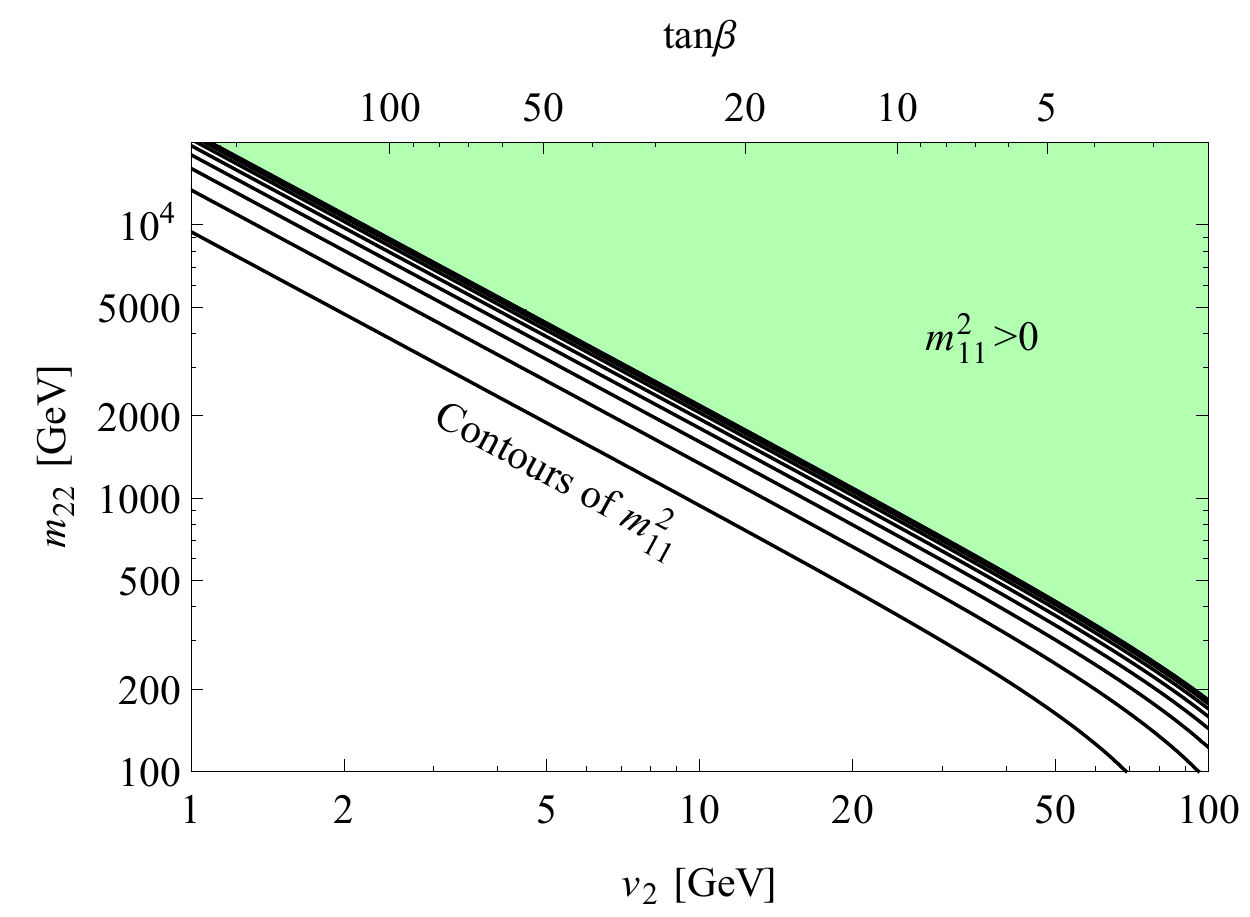}
 \caption{Contours of $m_{11}^2/\text{GeV}^2=-80^2,-70^2,$ and so on.
  The shaded region has no solution for $m_{11}^2<0$.}
 \label{Fig2HDMConsistency}
\end{figure}

There is a useful constraint on $m_{22}^2$ which is derived as follows.
The minimisation conditions give
\begin{align}
 m_{22}^2 + \frac12 \lambda_2 v_2^2 = \tan^2\beta \left( m_{11}^2 + \frac12 \lambda_1 v_1^2 \right), \label{EqMinCond}
\end{align}
where $\lambda_1 v_1^2\approx m_h^2 \approx (125\text{ GeV})^2$ (see below).
In the limit $m_{22}^2 \gg \lambda_2 v_2^2/2$, it can be seen that
$m_{22}^2$ is bounded above by 
\begin{align}
 m_{22}^2 \lesssim \frac12 m_h^2 \tan^2\beta \label{Eqconsistency}
\end{align}
if $m_{11}^2$ is to remain negative.
Figure~\ref{Fig2HDMConsistency} illustrates how $m_{11}^2$ deviates 
from its standard value of $-(88\text{ GeV})^2$
as $m_{22}^2$ approaches this bound.
For $m_{22}^2$ above this bound, $m_{11}^2$ very quickly grows to values $>v^2$,
and $v\approx 246$~GeV is only explained by a miraculous balance of $m_{11}^2$ against $m_{22}^2/\tan^2\beta$,
which constitutes a fine-tuning. 
Thus we adopt Eq.~\ref{Eqconsistency} as a consistency condition.

The charged scalar and pseudoscalar (neutral scalar) mass-squared matrices 
are diagonalised by a mixing angle $\beta$ ($\alpha$).
The neutral mass eigenstates are
\begin{align}
 h &= \rho_1\cos\alpha + \rho_2\sin\alpha  \nonumber,\\
 H &= \rho_2\cos\alpha - \rho_1\sin\alpha \nonumber, \\
 A &= \eta_2\sin\beta - \eta_1\cos\beta ,
\end{align}
where $\rho_i = \sqrt{2}\text{Re}(\Phi_i^0)-v_i$ and $\eta_i = \sqrt{2}\text{Im}(\Phi_i^0)$.
The masses are given by
\begin{align}
 m_h^2 		&= \lambda_1 v_1^2 + \mathcal{O}\left(\frac{m_{12}^4}{m_{22}^4}v_1^2\right), \nonumber \\
 m_H^2 		&= m_{22}^2 + \frac12 \lambda_{345}v_1^2 + \mathcal{O}\left(\frac{m_{12}^4}{m_{22}^4}m_{22}^2\right), \nonumber  \\
 m_A^2		&= m_{22}^2 + \frac12 (\lambda_{345}-2\lambda_5)v_1^2  + \mathcal{O}\left(\frac{m_{12}^4}{m_{22}^4}m_{22}^2\right), \nonumber  \\
 m_{H^\pm}^2	&= m_{22}^2 + \frac12 \lambda_3v_1^2  + \mathcal{O}\left(\frac{m_{12}^4}{m_{22}^4}m_{22}^2\right), \label{Eqscalarmasses}
\end{align}
i.e. the same as in the inert doublet model \cite{Branco2011iw} up to 
corrections proportional to $m_{12}^4/m_{22}^4$, which we provide in Appendix~\ref{AppMasses}.
Clearly, if $m_{22}^2\gg v^2$, the mass scale of extra scalar states is $\approx m_{22}$.

In the alignment limit $\cos(\alpha-\beta)\to 0$,
the couplings of $h$ to SM particles become SM-like.
We calculate
\begin{align}
 \cos^2&(\alpha-\beta) \approx \nonumber \\
 &\frac{m_{12}^4}{m_{22}^4}\frac{v_1^4}{m_{22}^4}
 \frac{\left(\lambda_1-\lambda_{345}\right)^2}
 { 
  \left(1-\frac{v_1^2}{2m_{22}^2}(2\lambda_1-\lambda_{345})\right)^2 
  \left(1+\frac{v_1^2}{2m_{22}^2}\lambda_{345}\right)^2 
  } ,\label{Eqcosbmasq}
\end{align}
suppressed by the approximate $U(1)$ or $Z_2$ symmetry ($m_{12}^2/m_{22}^2\ll 1$)
as well as the usual decoupling limit suppression ($v_1^2/m_{22}^2\ll 1$) \cite{Gunion2002zf}.
Thus the model naturally accommodates a SM-like neutral scalar state.

\subsection{Constraints \label{SecConstraints}}

With $M_N > m_{22}$ the constraints (and search strategies) for a $\nu$2HDM of given Type
are largely identical to those for a 2HDM of the same Type,
for which there is extensive literature (see references henceforth).
The 2HDM potential Eq.~\ref{EqV2HDM} is subject to a few standard theoretical constraints \cite{Branco2011iw}.
The necessary and sufficient conditions for 
positivity of the potential in all directions are \cite{Deshpande1977rw,Klimenko1984qx,Maniatis2006fs}
\begin{align}
 \lambda_{1,2} &\ge 0, \nonumber \\
 \lambda_3 &\ge -\sqrt{\lambda_1\lambda_2}, \nonumber \\
 \lambda_3+\lambda_4-|\lambda_5| &\ge -\sqrt{\lambda_1\lambda_2}.
\end{align}
Vacuum stability of the potential minimum is more difficult to evaluate.
An inequality which ensures a global minimum,
missing possible metastable vacua, is presented in Ref.~\cite{Barroso2013awa}.
Tree-level perturbative unitarity of scalar-scalar scattering is ensured by bounding
the eigenvalues of the scattering matrix \cite{Branco2011iw,Kanemura1993hm,Akeroyd2000wc,Ginzburg2005dt}.
Perturbativity of the $\lambda_i$ can also be demanded \cite{Nierste1995zx}.
At the very least these bounds should be implemented at the mass scale of the scalar states.
In addition they may be demanded up to some high scale under the renormalisation group evolution,
which results in non-trivial constraints on the parameter space (see e.g. Ref.~\cite{Chakrabarty2014aya,Das2015mwa,Chowdhury2015yja}\footnote{Note 
that some of the bounds derived in these papers do not apply to the softly broken $Z_2$-symmetric case, 
and also do not apply to the $Z_2$-symmetric case when one of the vevs vanishes}).
Type II, LS, and Flipped 2HDMs are particularly susceptible to exclusion by such a demand at large $\tan\beta$;
at one-loop, their Yukawa couplings hit a Landau pole before $M_{Pl}\sim 10^{18}$~GeV
when $v_2\lesssim 3.6,2.3,3.3$~GeV ($\tan\beta\gtrsim 68,107,75$) respectively \cite{Bijnens2011gd}.
These Landau poles merely indicate the breakdown of perturbativy.

The scalar boson discovered at the LHC is to be identified with the mass eigenstate $h$.
Its couplings have been measured to be SM-like, 
which constrains the $\nu$2HDM to lie in the alignment limit $\cos(\alpha-\beta)\approx 0$,
particularly at large $\tan\beta$, and for Type II and Flipped 2HDMs.\footnote{We 
refer the reader to 
Refs.~\cite{Chen2013rba,Eberhardt2013uba,Craig2013hca,Belanger2013xza,Chang2013ona,Dumont2014wha,Chowdhury2015yja} 
for allowed parameter space 
as a function of $\cos(\alpha-\beta)$ and $\tan\beta$ in all 2HDM Types.}
As is evident from Eq.~\ref{Eqcosbmasq}, the alignment limit is automatically preferred in our model
due to the approximate $Z_2$ or $U(1)$ symmetry.
Thus we limit the following discussion on additional 
experimental limits to those that constrain
moderate to large $\tan\beta$ models very close to the alignment limit.

In Type II and Flipped 2HDMs, the $\Phi_2$ coupling to down-type quarks is $\tan\beta$ enhanced.
The $H^\pm$ state then contributes significantly to radiative $B\to X_s \gamma$ decay;
the experimental measurement \cite{Amhis2014hma} 
combined with a recent next-to-next-to-leading order SM calculation \cite{Misiak2015xwa}
bounds $m_{H^\pm}\gtrsim 480$~GeV at 95\% CL for $\tan\beta\gtrsim 2$.
This bound along with the consistency condition Eq.~\ref{Eqconsistency}
implies $v_2\lesssim 45$~GeV ($\tan\beta\gtrsim 5.4$) for these $\nu$2HDMs.
In the Type II 2HDM the $\Phi_2$ coupling to $e_R^i$ is also $\tan\beta$ enhanced,
and the bound on $m_{H^\pm}$ from $B\to \tau\nu$ decays exceeds the radiative
bound when $\tan\beta\gtrsim 60$ \cite{Branco2011iw}.

Direct searches at LEP constrain $m_{H^\pm}\gtrsim 80$~GeV assuming decay to SM particles \cite{Searches2001ac}.
At the LHC, searches for $H/A\to \tau\tau$ \cite{Aad2014vgg,Khachatryan2014wca} are particularly
constraining in the Type II 2HDM.
The 95\% CL limit rises approximately linearly from 
$m_A\gtrsim 300$~GeV at $\tan\beta=10$ to $m_A\gtrsim 1000$~GeV at $\tan\beta=60$.
Such searches can also be mildly constraining for the LS 2HDM at moderate $\tan\beta$.
Searches for $H^\pm\to\tau\nu$ \cite{Aad2014kga,CMS2014cdp} 
cannot compete with $B\to X_s \gamma$ for Type II/Flipped 2HDMs
or with $H/A\to\tau\tau$ for the LS 2HDM.
However, for $m_{H^\pm}<160$~GeV, significant parameter space is ruled out
in Type I 2HDMs with moderate $\tan\beta$.

The $(y_\nu)_{ij} \overline{l_L^i}\tilde\Phi_2 \nu_R^j$ Yukawa term related to the neutrino masses 
can induce lepton flavour violating decays;
these are suppressed by the small $y_\nu$ and the right-handed neutrino mass scale $M_N > m_{22}$.
The processes of interest are $l_\alpha\to l_\beta\gamma$, $l_\alpha\to 3l_\beta$, and $\mu\to e$ conversion in nuclei
(see Ref.~\cite{Toma2013zsa} for expressions).
As well, $b\to s l_\alpha \bar{l}_\beta$ decays are induced in Type II and Flipped $\nu$2HDMs.
In practice, lepton flavour violating measurements 
constrain linear combinations of $(y_\nu)_{ij}$ bi- and tri-linears as well as the $M_{N_i}$.

In summary, for moderate to large $\tan\beta$ and $\cos(\alpha-\beta)\approx 0$, 
experiments are most constraining
for the Type II and Flipped 2HDMs, with $m_{22}\gtrsim 480$~GeV necessary
(implying $v_2\lesssim 45$~GeV).
For Type I and LS 2HDMs, even additional scalars with masses down to $80$~GeV
may still have evaded detection.

\section{Naturalness \label{Secnu2hdmNat}}

In the SM, the renormalisation group equation (RGE) for the electroweak $\mu$ parameter 
(as in Eq.~\ref{EqVSM}) is dominated by the top quark Yukawa,
\begin{align}
 \frac{d\mu^2}{d\ln\mu_R} \approx \frac{1}{(4\pi)^2} 6 y_t^2 \mu^2 ,
\end{align}
where $\mu_R$ is the renormalisation scale.
At low energy we measure $\mu^2 \approx -(88\text{ GeV})^2$, 
and under SM running it is apparent that $|\mu|$ remains $\sim 100$~GeV 
even up to the Planck scale $M_{Pl}\sim10^{18}$~GeV.
Thus there is no \textit{measurable} naturalness problem in the SM alone;
there is no fine-tuning of any measurable parameter at a high scale,
only the cancellation of an unmeasurable bare parameter against an unphysical cutoff scale,
which should be assigned no physical significance.
With this understood, it is clear that a measurable naturalness problem 
can only arise when $d\mu^2/d\ln\mu_R\gtrsim \mu^2$.
Indeed, this is exactly how the Vissani bound in the Type I see-saw
model can be interpreted \cite{Vissani1997ys,Clarke2015gwa}.
Let us now examine when the $\nu$2HDM encounters such a problem.

In practice, the naturalness considerations can be divided into two distinct calculations: 
the influence of $m_{22}$ on $m_{11}$, and the influence of $M_{N}$ on $m_{22}$.
These influences will be considered in turn.\footnote{In
the following we ignore the influence of the small $y_\nu$ Yukawas on $m_{11}^2$,
and hence those results also hold in a general 2HDM.}

\subsection{Corrections to $m_{11}^2$}

If $m_{22}^2 \ll m_h^2\tan^2\beta/2$, then $m_{11}^2$
sets the mass of the observed SM-like Higgs via Eqs.~\ref{Eqvev1} and \ref{Eqscalarmasses}.
The one-loop RGE for the $m_{11}^2$ parameter
is \cite{Branco2011iw} (see Ref.~\cite{Chowdhury2015yja} for a recent two-loop calculation)
\begin{align}
 \frac{dm_{11}^2}{d\ln\mu_R} = \frac{1}{(4\pi)^2} \left[ (4\lambda_3+2\lambda_4)m_{22}^2 + \mathcal{O}(m_{11}^2) \right]. \label{Eqm11sqRGE}
\end{align}
The $\mathcal{O}(m_{11}^2)$ term contains gauge, $\lambda_1$, and Yukawa contributions,
which, as in the SM case, do not induce a naturalness problem.
However if $\lambda_{3,4}$ are non-zero then a naturalness problem is induced for sufficiently large $m_{22}^2$;
we are interested in when this generically occurs.
Even if $\lambda_{3,4}=0$ at some scale,
they will quickly be reintroduced by gauge interactions at one-loop.
Their one-loop RGEs are given by
\begin{align}
 \frac{d\lambda_3}{d\ln\mu_R} &= \frac{1}{(4\pi)^2} \left[ \frac34\left(g_Y^4-2g_Y^2g_2^2+3g_2^4\right) + ... \right],  \nonumber \\
 \frac{d\lambda_4}{d\ln\mu_R} &= \frac{1}{(4\pi)^2} \left[ 3g_Y^2g_2^2 + ... \right], \label{Eqlam34RGEs}
\end{align}
where $g_2^2(m_Z)\approx 0.43$ and $g_Y^2(m_Z)\approx 0.13$ are the gauge couplings
and the ellipses contain terms multiplicative in $\lambda_{3,4}$, terms proportional to $\lambda_5^2$,
and terms related to the Yukawas.
Let us ignore those effects for now and return to them later.
Note that ignoring the contribution from $\lambda_5^2$
is equivalent to assuming $\lambda_5 \lesssim 0.2$, 
so that its contribution is subdominant to the gauge couplings.
Typically, one would expect
\begin{align}
 |\lambda_3(\mu_R)| &\gtrsim \frac{1}{(4\pi)^2} \frac34\left(g_Y^4-2g_Y^2g_2^2+3g_2^4\right) , \nonumber \\
 |\lambda_4(\mu_R)| &\gtrsim \frac{1}{(4\pi)^2} 3g_Y^2g_2^2 ,
\end{align}
and thus
\begin{align}
 \left| \frac{dm_{11}^2}{d\ln\mu_R} \right| \gtrsim \frac{1}{(4\pi)^4} \left(3g_Y^4+9g_2^4\right)m_{22}^2 .\label{Eqm11RGEmin}
\end{align}
This lower bound is of the same order as the two-loop pure gauge contribution \cite{Chowdhury2015yja}.

Equation~\ref{Eqm11RGEmin} represents a conservative bound on the running of the $m_{11}^2$ parameter
above the scale $\sim m_{22}$.
Naturalness demands that this running not be significantly larger than 
the value measured at a low scale, $|m_{11}|\approx 88$~GeV.
A very conservative naturalness bound is therefore
\begin{align}
 \frac{1}{(4\pi)^4} \left(3g_Y^4+9g_2^4\right)m_{22}^2 < 1\text{ TeV}^2,\\
 \Rightarrow m_{22} \lesssim 1\times 10^5\text{ GeV}. \label{EqNatBound1}
\end{align}

Alternatively, we can try to bound a quantity which 
measures the fine-tuning in $m_{11}^2$ at some high scale $\Lambda_{h}$.
A typical quantity is \cite{Ellis1986yg,Barbieri1987fn}
\begin{align}
 \Delta\left(\Lambda_h\right) = \left| \frac{m_{11}^2(\Lambda_h)}{m_{11}^2(0)}
   \frac{\partial m_{11}^2(0)}{\partial m_{11}^2(\Lambda_h)} \right|, \label{EqDelta}
\end{align}
which compares percentage variations of two (in principle) measurable parameters.
Let us now estimate how such a bound might constrain $m_{22}$.

For simplicity, and anticipating that the $m_{22}$ scale is not far above the
electroweak scale, we will evolve the dimensionless parameters using the ($\nu$)2HDM RGEs from the $m_Z$ scale.
First, the one-loop gauge coupling RGEs \cite{Grzadkowski1987wr,Branco2011iw} can be solved analytically.
Upon substitution into the $\lambda_{3,4}$ RGEs (Eqs.~\ref{Eqlam34RGEs}),
and considering only the pure gauge contribution,
the $\lambda_{3,4}$ running can be solved for given initial conditions.
For simplicity we take $\lambda_3(m_{22})=\lambda_4(m_{22})\equiv \lambda_{3,4}(m_{22})$ and consider it a free parameter.
Next we solve Eq.~\ref{Eqm11sqRGE} for $m_{11}^2(\mu_R)$ with 
the initial condition $m_{11}^2(m_{22})= -(88$~GeV$)^2$
(neglecting any RGE evolution of $m_{22}^2$).
With these simplifications $\partial m_{11}^2(0)/\partial m_{11}^2(\Lambda_h)=1$
and the fine-tuning measure is given simply by 
$\Delta(\Lambda_h)=\left| m_{11}^2(\Lambda_h)/(88\text{ GeV})^2 \right|$.

Note that in setting the initial condition $m_{11}^2(m_{22})= -(88$~GeV$)^2$
we have implicitly assumed that $m_{22}^2 \ll m_h^2\tan^2\beta /2$
(see Eq.~\ref{Eqconsistency} and Fig.~\ref{Fig2HDMConsistency}).
This is conservative for negative $m_{11}^2$,
since $|m_{11}^2(m_{22})|$ shrinks as $m_{22}^2/\tan^2\beta \to m_h^2/2$
and the naturalness constraint would become more stringent.
In some circumstances we will obtain naturalness bounds on $m_{22}^2$ which exceed $m_h^2\tan^2\beta /2$,
which just indicates that the naturalness constraint is weaker than
the consistency condition Eq.~\ref{Eqconsistency}.

\begin{figure}[t]
 \centering
 \includegraphics[width=0.9\columnwidth]{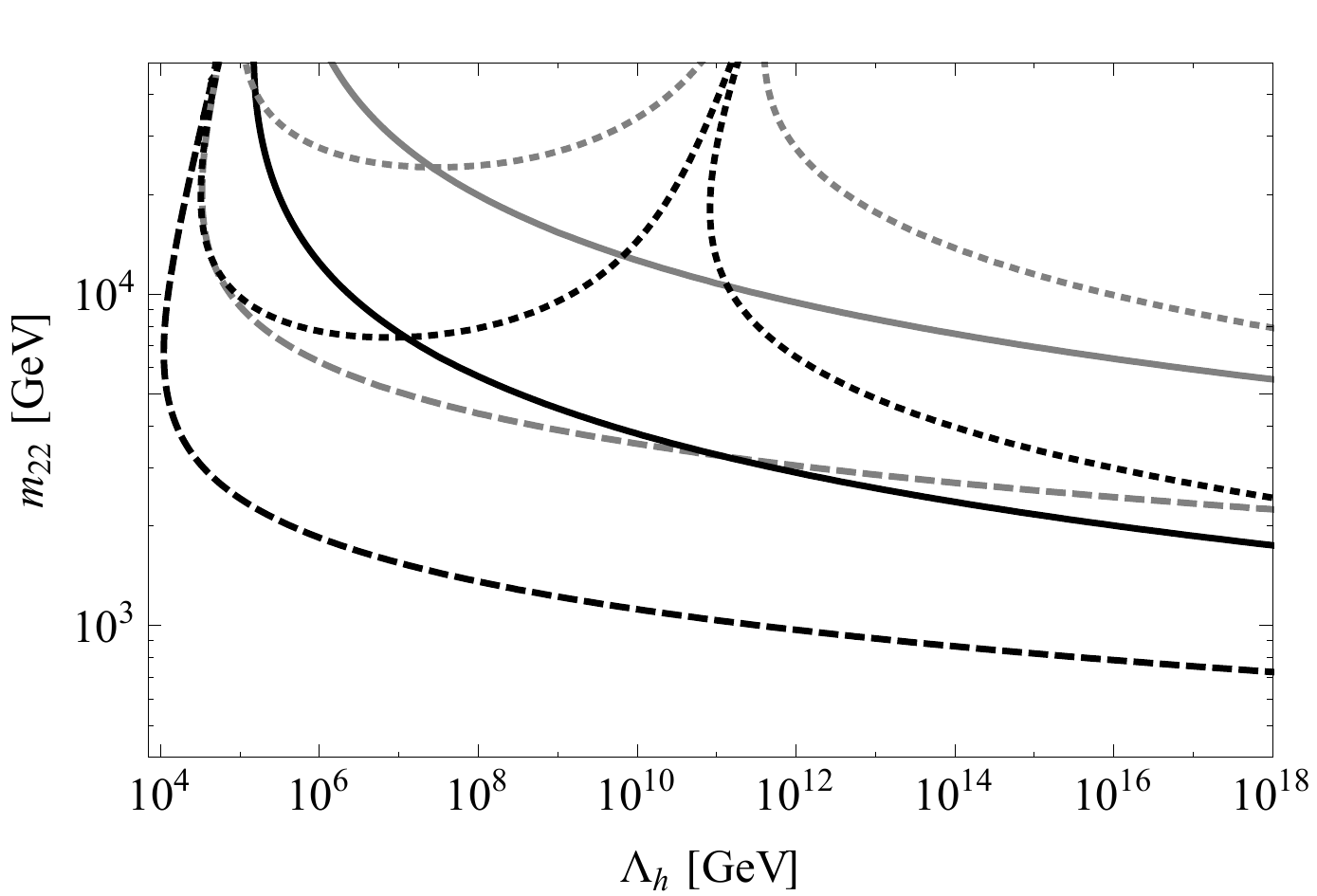}\\
 \includegraphics[width=0.9\columnwidth]{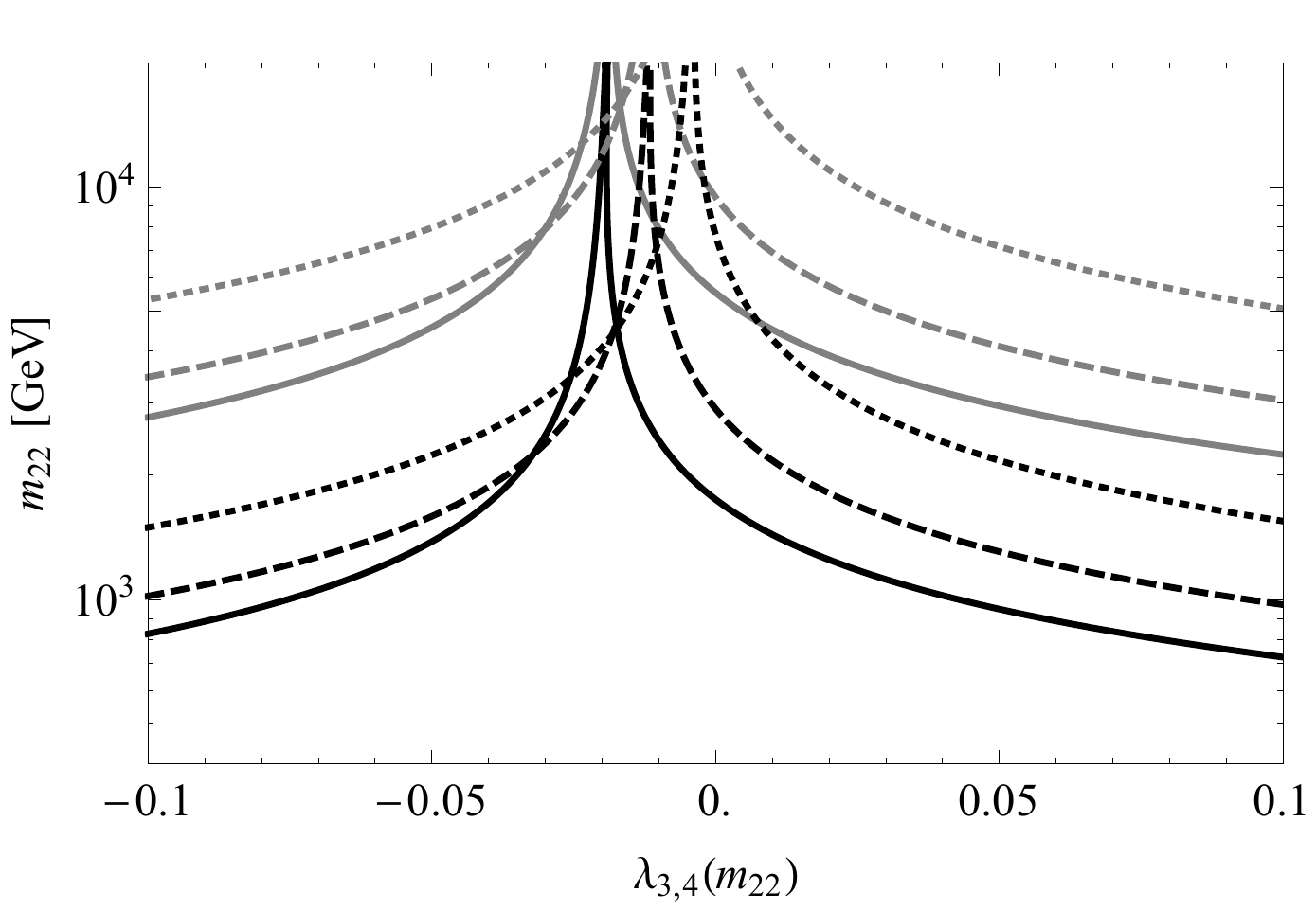}
 \caption{Contours of the fine-tuning measure $\Delta(\Lambda_h)=10$ (100) in black (grey)
  as a function of (top) $\Lambda_h$ for $\lambda_{3,4}(m_{22})=0.0,0.1,-0.01$ (solid, dashed, dotted),
  and (bottom) $\lambda_{3,4}(m_{22})$ for $\Lambda_h=10^{18},10^{12},10^7$~GeV (solid, dashed, dotted).
  See the text for the assumptions that accompany this plot.}
 \label{FigNatural2HDM}
\end{figure}

In Fig.~\ref{FigNatural2HDM} we show $\Delta=10$ and $\Delta=100$ contours
as a function of $\Lambda_h$ and $\lambda_{3,4}(m_{22})$.
These represent naturalness upper bounds on $m_{22}$.
The cusp-like structures of apparently low fine-tuning in $m_{11}^2$ 
occur when $m_{11}^2$ runs negative before turning and passing through $m_{11}^2=0$, 
which just corresponds to a fine-tuning in $(\lambda_{3,4},\Lambda_h)$.
A stringent naturalness constraint is obtained by 
demanding $\Delta<10$ at $\Lambda_h = M_{Pl}$;
from Fig.~\ref{FigNatural2HDM}, it is clear that this implies
\begin{align}
 m_{22}\lesssim 
  \text{few}\times 10^3\text{ GeV}.
 \label{EqNatBound2}
\end{align}
If any new physics comes in below $M_{Pl}$ then the running of $m_{11}^2$ could change,
and these bounds do not apply.
If that is the case then it is more appropriate to consider $\Lambda_h$ at the scale of the new physics,
which weakens the bound, as is clear from Fig.~\ref{FigNatural2HDM}.
In the $\nu$2HDM this new physics scale is the right-handed neutrino scale $M_N$,
after which the right-handed neutrinos can
contribute to the running of $m_{11}^2$ through $m_{22}^2$ at one-loop.

\begin{figure}[t]
 \centering
 \includegraphics[width=0.9\columnwidth]{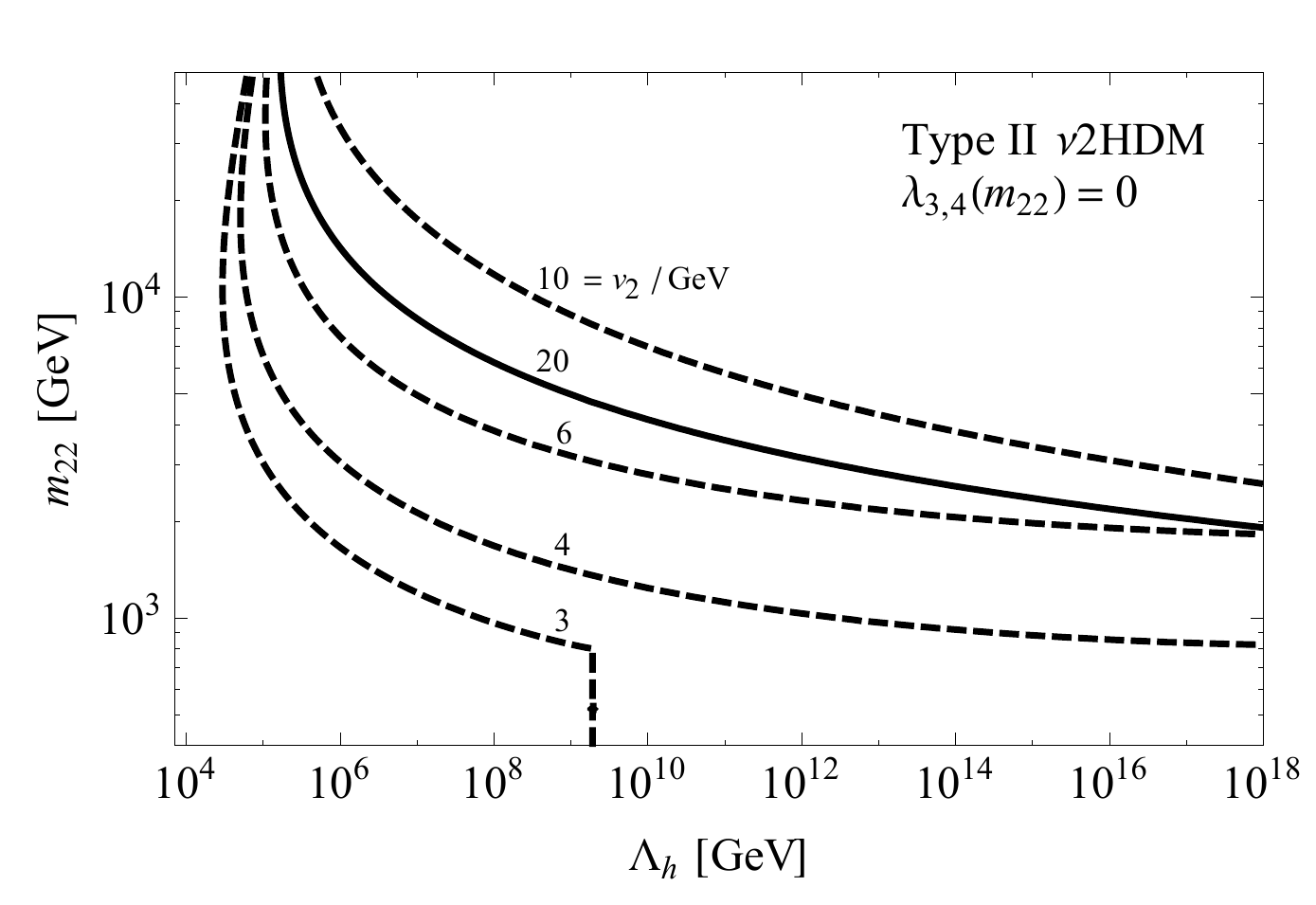}\\
 \includegraphics[width=0.9\columnwidth]{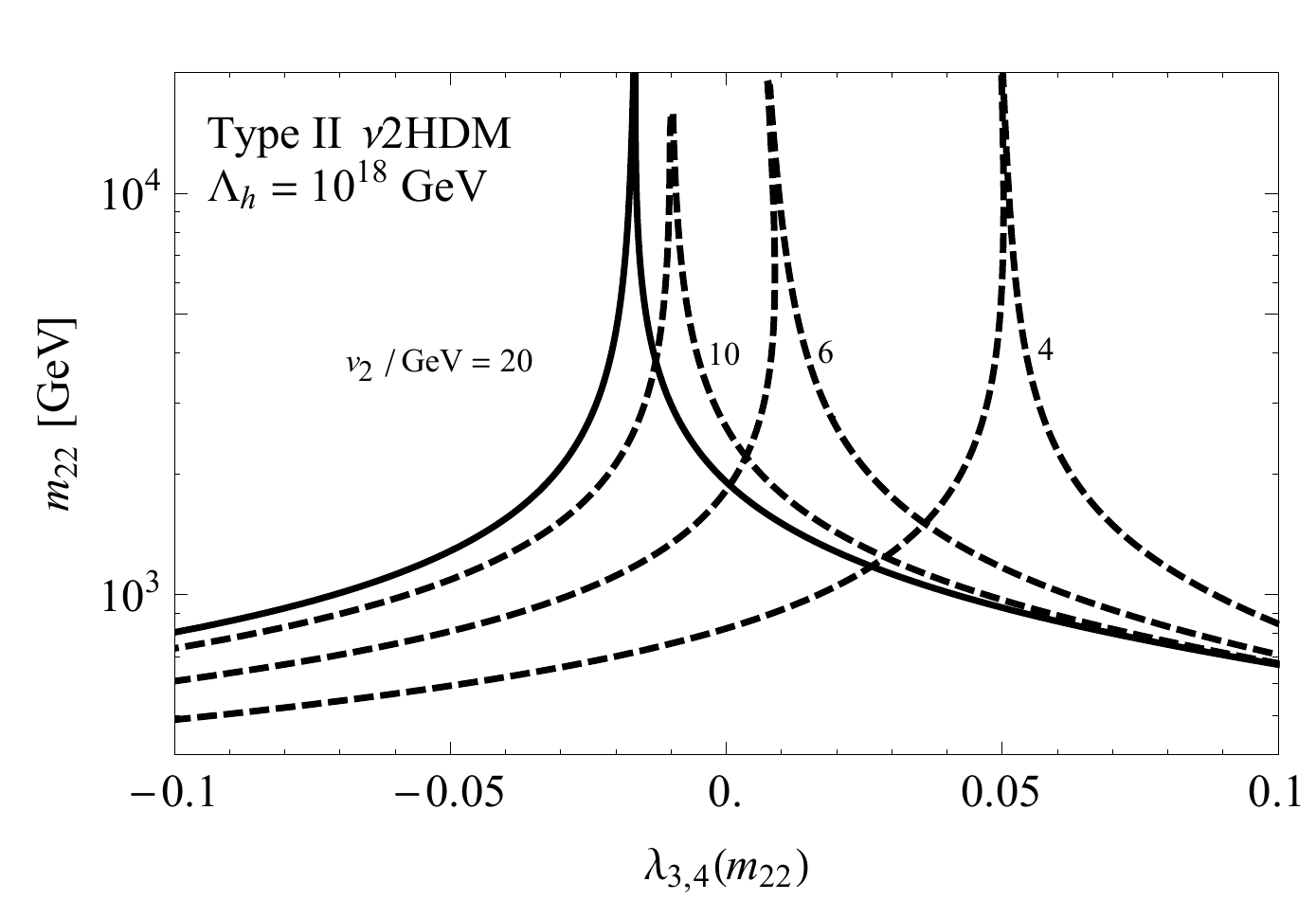}
 \caption{Contours of the fine-tuning measure $\Delta(\Lambda_h)=10$ 
  in an illustrative Type II $\nu$2HDM (see text) for different values of $v_2$.
  The solid lines for $v_2=20$~GeV match approximately onto the solid lines in Fig.~\ref{FigNatural2HDM}.}
 \label{FigNatural2HDMT2}
\end{figure}

We have so far ignored the RGE contributions from possibly large Yukawas.
There are two situations in which the Yukawas play a significant role.
The first is in Type II, LS, and Flipped $\nu$2HDMs
with $\tan\beta$ large enough such that an early Landau pole is induced (see Sec.~\ref{SecConstraints}),
and the second is in Type II and Flipped $\nu$2HDMs with moderate to large $\tan\beta$
when the pure Yukawa term $\pm \frac{1}{(4\pi)^2}12y_b^2y_t^2$ induced by a quark box diagram
contributes significantly to the $\lambda_{3,4}$ RGEs (Eqs.~\ref{Eqlam34RGEs}).
In Fig.~\ref{FigNatural2HDMT2} we show how the $\Delta=10$ contours change as a function of $v_2$
in an example Type II $\nu$2HDM.
For this Figure we have numerically solved the full set of one-loop RGEs \cite{Chowdhury2015yja} 
including the top/bottom/tau Yukawas, 
taking the following values at the scale $M_Z$: $g_s^2=1.48$, $\lambda_1=\lambda_2=0.26$,
and $y_t=0.96/\sin\beta$, $y_b=0.017/\cos\beta$, $y_\tau=0.010/\cos\beta$.
Comparing to Fig.~\ref{FigNatural2HDM}, 
it can be seen that the pure Yukawa term has a noticeable effect when $v_2\lesssim 20$~GeV.
It is also apparent from Fig.~\ref{FigNatural2HDMT2}
that nearing $v_2\approx 3.6$~GeV (below which a Landau pole is induced before $M_{Pl}$)
can act to degrade \textit{or} improve the naturalness bound.
The $v_2=3$~GeV bound in Fig.~\ref{FigNatural2HDMT2} shows the effect of hitting the Landau pole at $\sim 10^9$~GeV.
We note that this only signals the breakdown of perturbation theory, and of our one-loop RGEs;
we cannot calculate $m_{22}(\mu_R)$ above this scale
though it is perfectly possible that the theory remains natural.

In a repeated full one-loop RGE analysis we found that the Flipped $\nu$2HDM 
gave essentially the same results as the Type II $\nu$2HDM in Fig.~\ref{FigNatural2HDMT2},
and there was no noticeable Yukawa effect in the LS $\nu$2HDM until the Landau pole was reached.
Thus we find that the stringent naturalness bounds of Eq.~\ref{EqNatBound2} and Fig.~\ref{FigNatural2HDM} 
are applicable at all times in the Type I $\nu$2HDM, 
for $v_2\gtrsim 2$~GeV in the LS $\nu$2HDM, and for $v_2\gtrsim 20$~GeV in the Type II and Flipped $\nu$2HDMs.
Otherwise Yukawa effects must be taken into account.
Either way, the important point is now clear: a TeV-scale $m_{22}$ can be both completely natural
and, as was discussed in the previous subsection, is experimentally allowed in all $\nu$2HDM Types.

\subsection{Corrections to $m_{22}^2$}

\begin{figure}[t]
 \centering
 \includegraphics[width=0.95\columnwidth]{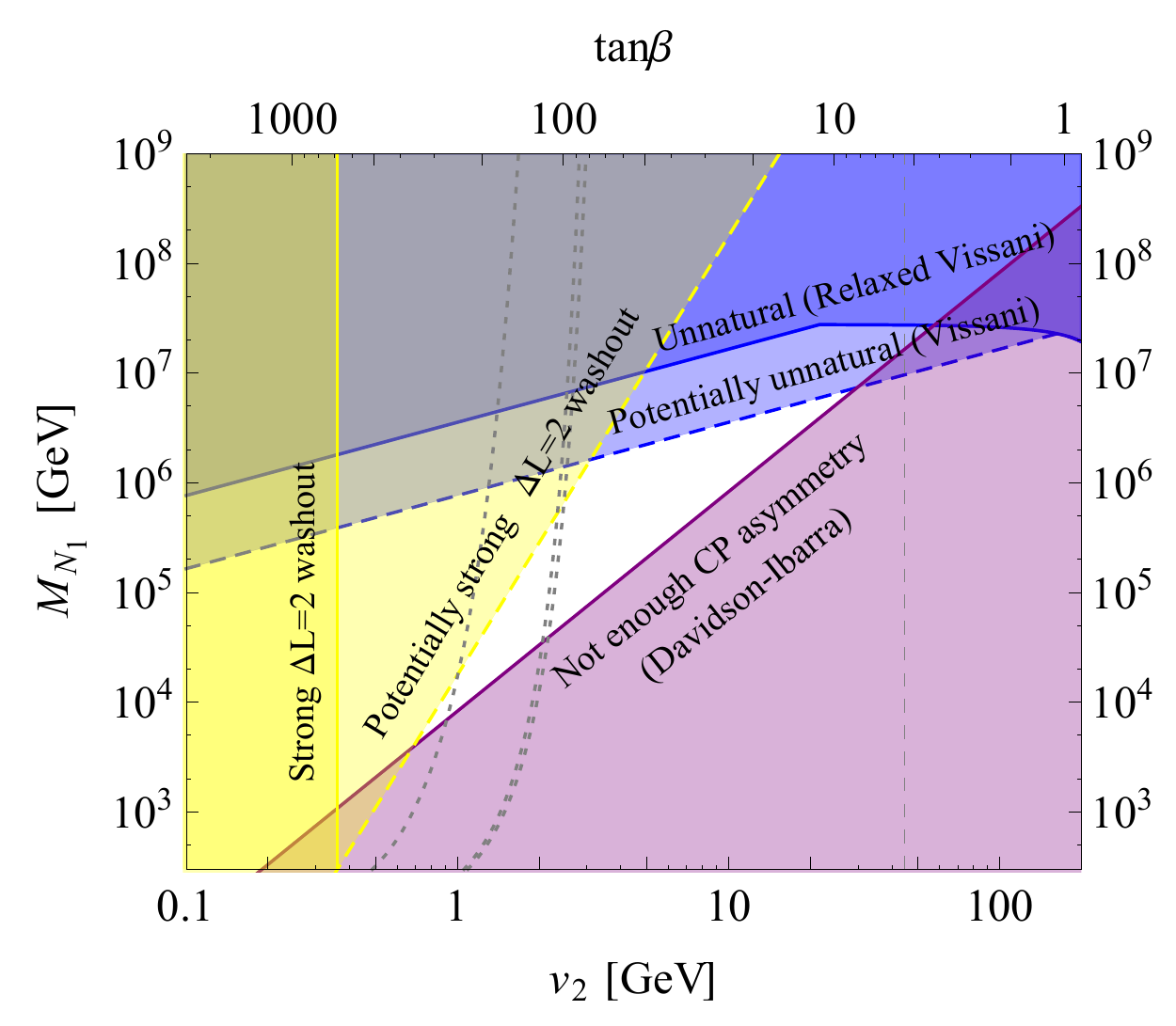}
 \caption{Bounds on the $\nu$2HDM as a function of $v_2$.
 Shown (as labelled) are the Davidson-Ibarra bound,
 the Vissani and Relaxed Vissani naturalness bounds, 
 and the areas of parameter space with strong $\Delta L=2$ scattering washout.
 The Type II and Flipped $\nu$2HDMs are excluded by $B\to X_s\gamma$
 for values of $v_2$ greater than indicated by the grey dashed line (see Sec.~\ref{SecConstraints}).
 The grey dotted lines indicate the $v_2$ below which 
 the Yukawas hit a Landau pole before $M_{N_1}$
 in the Type II, Flipped, and LS $\nu$2HDMs right-to-left.}
 \label{FigMNv2}
\end{figure}

Let us now consider the influence of the right-handed neutrinos.
The one-loop RGE for $m_{22}^2$ is \cite{Clarke2015gwa,Merle2015gea}
\begin{align}
  \frac{dm_{22}^2}{d\ln\mu_R} = \frac{1}{(4\pi)^2} 
  \left[ -4 \text{Tr}[y_\nu\mathcal{D}_M^2 y_\nu^\dagger] + \mathcal{O}(m_{22}^2) \right]. \label{Eqm22sqRGE}
\end{align}
A conservative naturalness bound is obtained by bounding the running as we did in Eq.~\ref{EqNatBound1},
\begin{align}
 \frac{1}{4\pi^2} \text{Tr}[y_\nu\mathcal{D}_M^2 y_\nu^\dagger] < \Lambda_{bound}^2, \label{EqNatBoundMN}
\end{align}
where taking $\Lambda_{bound}=1$~TeV gives the Vissani bound on $M_{N_1}$ of Eq.~\ref{EqNatural} \cite{Clarke2015gwa}.
However, now we are bounding corrections to $m_{22}^2$ rather than $m_{11}^2$, 
which may be TeV-scale.
Thus, depending on the mass of the extra scalars, 
it is possible that we can sensibly take $\Lambda_{bound}>1$~TeV,
in which case the the naturalness bound is somewhat relaxed;
in Fig.~\ref{FigMNv2} we show a Relaxed Vissani bound for 
$\Lambda_{bound}=\min (10\text{ TeV},10 \sqrt{m_h^2\tan^2\beta/2})$,
where we have kept in mind the consistency condition Eq.~\ref{Eqconsistency}.
The Vissani bound still represents the unnatural area of parameter space if $m_{22}$ is closer to 100~GeV.

As before, we could instead bound a quantity which 
measures the fine-tuning in $m_{22}^2$ at some high scale $\Lambda_{h}$.
In this case, the fine-tuning measure of Eq.~\ref{EqDelta} is
\begin{align}
 \Delta(\Lambda_h) = 1 + \frac{1}{4\pi^2}\frac{\sum_{i,j} (y_\nu)_{ij} M_j^2 (y_\nu^\dagger)_{ji} \ln(M_j/\Lambda_h)}{m_{22}^2(M_j)}.
\end{align}
Taking $m_{22}(M_j)\sim 1$~TeV and demanding $\Delta(M_{Pl})<10$ gives a similar bound to Vissani
(Eq.~\ref{EqNatBoundMN} with $\Lambda_{bound}=1$~TeV).
Note that there is no naturalness bound on $M_N$ in the $y_\nu\to 0$ limit.
This is the technically natural limit corresponding to an enhanced Poincar\'e symmetry
in which $\nu_R$ decouples from the theory \cite{Foot2013hna}.

In summary, there are up to three scales in the $\nu$2HDM: $v$, $m_{22}$, and $M_N$.
We have described the conditions under which $v^2$ (or $m_{11}^2$) is protected from $m_{22}^2$, 
and $m_{22}^2$ from $M_N^2$.
Under such conditions it follows that $m_{11}^2$ is also protected from $M_{N_1}^2$
and the model is entirely natural.

\begin{figure}[t]
 \centering
 \includegraphics[width=0.95\columnwidth]{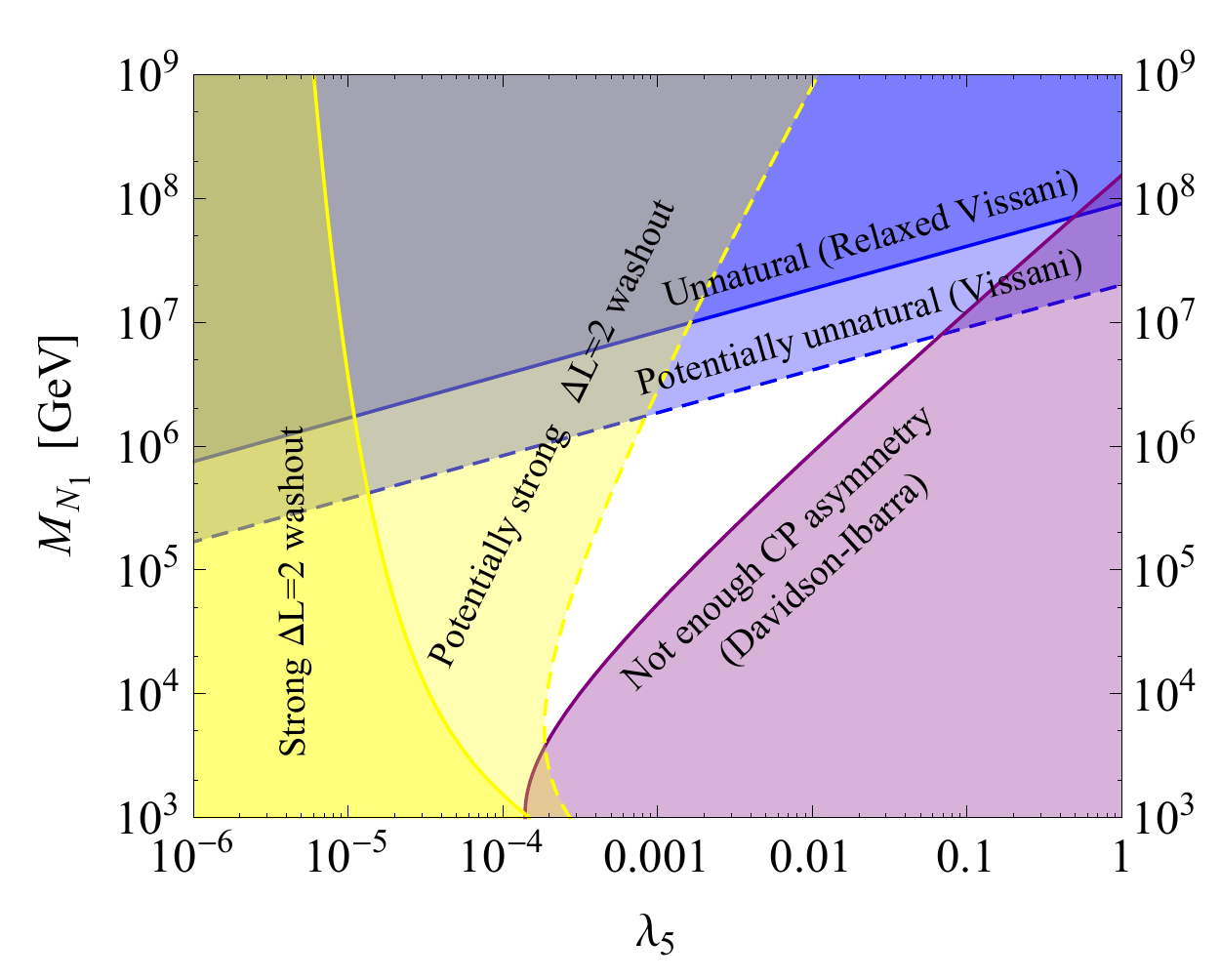}
 \caption{As in Fig.~\ref{FigMNv} but for the Ma model. 
 The Vissani and Relaxed Vissani bounds are evaluated at $m_{22}=100,1000$~GeV respectively.
 The Davidson-Ibarra bound and strong $\Delta L=2$ washout region are shown for $m_{22}=500$~GeV,
 though they are only mildly sensitive to $m_{22}$.}
 \label{FigMNlam5}
\end{figure}

\section{Neutrino masses \label{Secnumasses}}

If $m_{12}^2=0$ and $\lambda_5=0$ then a $U(1)$ lepton number symmetry
can be defined and neutrinos remain massless.
Let us now consider turning each non-zero in turn.

\subsection{$m_{12}^2 > 0,\ \lambda_5= 0$}

If $m_{12}^2>0$ then the $U(1)$ lepton number symmetry is softly broken,
i.e. the breaking does not force us to insert a non-zero $\lambda_5$ term
in order to introduce a divergent counterterm, and it is consistent to consider $m_{12}^2>0,\ \lambda_5=0$.

In this situation the neutrino mass matrix is given by the see-saw formula,
\begin{align}
 m_\nu = \frac{v_2^2}{2} y_\nu \mathcal{D}_M^{-1} y_\nu^T 
   \approx \frac{1}{\tan^2\beta} \frac{v^2}{2} y_\nu \mathcal{D}_M^{-1} y_\nu^T, \label{Eqnumass2HDM}
\end{align}
where $\tan\beta \approx m_{22}^2/m_{12}^2$ for $m_{22}^2\gg \lambda_{345}v_1^2$ (see Eq.~\ref{Eqvev2}).

The analogous Davidson-Ibarra and Vissani bounds are given by 
the standard Eqs.~\ref{EqDavIbarra} and \ref{EqNatural} with the replacement $v\to v_2$.
These bounds are depicted in Fig.~\ref{FigMNv2}.
If $v_2\lesssim 30$~GeV ($\tan\beta\gtrsim 8$) then both bounds are satisfied.
As well, as discussed in Sec.~\ref{Secnu2hdmNat}, if $m_{22}$ is TeV-scale the Vissani bound can be relaxed and
the required CP asymmetry needed to reproduce the BAU via leptogenesis 
may be naturally achieved for $v_2\lesssim 60$~GeV ($\tan\beta\gtrsim 4$).
In the Type I $\nu$2HDM, $v_2$ can be naturally $\ll$~GeV.
Otherwise, requiring a perturbative theory up to $M_{N_1}$
restricts $v_2\gtrsim 1$~GeV ($v_2\gtrsim 2$~GeV) for the 
LS (Type II/Flipped) $\nu$2HDM in the parameter space region of interest,
as depicted in Fig.~\ref{FigMNv2}.

\subsection{$m_{12}^2=0,\ \lambda_5\ne 0$}

In this situation $v_2=0$ and a $Z_2$ symmetry remains unbroken;
this is the scenario of Ma \cite{Ma2006km}.
The model yields a radiative neutrino mass and a dark matter candidate.
This is only possible in the Type I $\nu$2HDM,
since in any other Type the unbroken $Z_2$
forbids a Dirac mass term for any charged fermion coupling to $\Phi_2$.
Note that the limit $\lambda_5\to 0$ is technically natural,
since in that limit the $U(1)$ lepton number symmetry is reinstated.

If $M_N^2\gg m_{22}^2,v^2$
the radiatively induced neutrino mass matrix is
\begin{align}
 (m_\nu)_{ij} \approx \frac{v^2}{2} \frac{(y_\nu)_{ik} (y_\nu^T)_{kj}}{M_k} \frac{\lambda_5}{8\pi^2} 
 \left( \ln\left[ \frac{2M_k^2}{(m_H^2+m_A^2)}\right] -1 \right). \label{EqnumassMa}
\end{align} 
The analogous Davidson-Ibarra and Vissani bounds are given by 
the standard Eqs.~\ref{EqDavIbarra} and \ref{EqNatural} with the intuitive replacement 
$v^2\to v^2 \frac{\lambda_5}{8\pi^2} \left( \ln\left[ 2 M_{N_1}^2 / (m_H^2+m_A^2) \right] -1 \right)$.
This assumes that there is no fine-tuning in the complex $y_\nu$ parameters to reproduce the observed neutrino masses
(see Appendix~\ref{AppBounds} for details).
These bounds are depicted in Fig.~\ref{FigMNlam5}, 
where the Davidson-Ibarra bound has been evaluated at $m_{22}=500$~GeV as an illustrative example
(the bound is only mildly sensitive to $m_{22}$).
We find that the Ma model with $\lambda_5\lesssim 0.5$
can naturally achieve the required CP asymmetry to reproduce the BAU via hierarchical leptogenesis.\footnote{A 
similar observation was made in a recent paper \cite{Davoudiasl2014pya}.}

\subsection{$m_{12}^2 > 0,\ \lambda_5 \ne 0$}

In this case both the tree-level see-saw and the radiative
mechanism will contribute to the neutrino mass.
Both contributions are calculable, and either might dominate.
Note that it is still technically natural to
take $\lambda_5\to 0$ in this case, 
since it restores a softly broken $U(1)$ symmetry.
In other words, the $\lambda_5$ RGEs to all orders will be multiplicative in $\lambda_5$,
indicative of the fact that the soft-breaking term
can only generate finite $U(1)$-breaking corrections.

\section{Leptogenesis \label{SecLeptogen}}

The observed BAU is achieved analogously to standard hierarchical thermal leptogenesis \cite{Fukugita1986hr};
the out-of-equilibrium CP-violating decays of the lightest right-handed neutrino $N_1\to l\Phi_2$
create a lepton asymmetry which is transferred to the baryons by the electroweak sphalerons above $T\sim 100$~GeV.

The details of the leptogenesis are largely defined by the decay parameter
\begin{align}
 K = \frac{\Gamma_D}{H|_{T=M_1}} = \frac{\tilde m_1}{m_*},
\end{align}
comparing the rate for decays and inverse decays to the expansion rate
at the time of departure from thermal equilibrium.
Here, the rates
\begin{align}
 \Gamma_D & = \frac{1}{8\pi} (y_\nu^\dagger y_\nu)_{11} M_1, \\
 H & \approx \frac{ 17 T^2 }{M_{Pl}},
\end{align}
are typically rescaled and expressed in terms of an
effective neutrino mass $\tilde m_1$ and an equilibrium neutrino mass $m_*$,
\begin{align}
 \tilde m_1 & = \frac{ (y_\nu^\dagger y_\nu)_{11} v^2}{2 M_1}, \nonumber \\
 m_* & \approx 1.1 \times 10^{-3} \text{ eV} \left( \frac{v}{246\text{ GeV}} \right)^2 , \label{Eqmtildemstar}
\end{align}
where $v$ is the vev that enters the see-saw Eq.~\ref{EqSeesawYuk}.
In the $\nu$2HDM with $\lambda_5=0$ (with $m_{12}^2=0, \lambda_5\ne 0$) the analogous definitions make the replacement $v^2\to v_2^2$
($v^2\to v^2 \frac{\lambda_5}{8\pi^2} \left( \ln\left[ 2 M_{N_1}^2 / (m_H^2+m_A^2) \right] -1 \right)$).
Note that for the scenarios we are interested in (e.g. $v_2\ll v$),
$m_*$ is smaller than its usual value in standard leptogenesis.

When only decays and inverse decays are considered,
leptogenesis for given $K$ proceeds exactly as in standard hierarchical thermal leptogenesis 
(see e.g. Ref.~\cite{Buchmuller2004nz} for a review).
In the weak washout regime $K\ll 1$, the baryon asymmetry
strongly depends on the initial asymmetry and the initial $N_1$ abundance, 
with $N_1$ decays occuring at $T\ll M_1$.
The strong washout regime $K\gg 1$ is independent of the initial conditions,
and the asymmetry is generated as the $N_1$ fall out of thermal equilibrium.

The $2\leftrightarrow 2$ scatterings with $\Delta L = 1$ (see e.g. Ref.~\cite{Davidson2008bu})
provide a correction to the simple decays plus inverse decays picture;
they act to increase $N_1$ production at $T>M_1$ and contribute to washout at $T<M_1$.
In standard hierarchical thermal leptogenesis, 
the scattering contributions involving the top quark
and the gauge bosons are roughly equal.
In the present model the gauge boson contribution is the same as in the standard scenario.
However, by construction, the $\Phi_2$ involved here in leptogenesis does not couple directly to the top quark,
and thus the usual s-channel ($Nl \leftrightarrow tq$) 
and t-channel ($Nt \leftrightarrow lq, Nq \leftrightarrow lt$) scattering contributions do not occur.
Instead, at large $\tan\beta$ they can be replaced by the analogous contribution from other charged fermions,
i.e. the bottom quark in Type II and Flipped $\nu$2HDMs
and/or the tau lepton in Type II and LS $\nu$2HDMs.
A large tau lepton Yukawa will also introduce new s-channel ($N\Phi_2 \leftrightarrow \tau\Phi_2$)
and t-channel ($N\Phi_2 \leftrightarrow \tau\Phi_2, \tau N \leftrightarrow \Phi_2\bar{\Phi}_2$)
scattering contributions.
All of these processes are proportional to $(y_\nu^\dagger y_\nu)_{11}$ and hence $M_1 \tilde m_1 / v^2$,
with the appropriate $\nu$2HDM replacement for $v^2$.
Therefore, they scale with the decays and inverse decays
so that they represent only a minor (but obviously important)
departure from the standard leptogenesis scenario.

The $2\leftrightarrow 2$ scatterings with $\Delta L = 2$ mediated by the right-handed neutrinos
($\Phi_2 l \leftrightarrow \bar{\Phi}_2\bar{l}, \Phi_2\Phi_2 \leftrightarrow ll$)
occur as they do in the standard scenario.
These processes are proportional to Tr$[(y_\nu y_\nu^T) (y_\nu y_\nu^T)^\dagger]$ 
and hence $M_1^2 \overline{m}^2 / v^4$ where $\overline{m}^2=\sum m_i^2$ 
is the neutrino mass scale $\gtrsim 0.05$~eV.
Comparing this rate to the decay/scattering rates $\propto M_1 \tilde m_1 / v^2$,
it is easy to see that after making the appropriate $\nu$2HDM replacement for $v^2$,
e.g. $v^2\to v_2^2\ll v^2$, these scatterings will become comparatively more important than in the standard case.
For $T\lesssim M_1/3$ the thermally averaged $\Delta L = 2$ scattering rate is well approximated by \cite{Buchmuller2004nz}
\begin{align}
 \frac{\Gamma_{\Delta L=2}}{H} \approx \frac{T}{2.2\times 10^{13}\text{ GeV}}
  \left(\frac{246\text{ GeV}}{v}\right)^4 \left(\frac{\overline{m}}{0.05\text{ eV}}\right)^2 , \label{EqDL2Scat}
\end{align}
where the previously described $\nu$2HDM replacements for $v^2$ hold (see Appendix \ref{AppBounds}).
In Fig.~\ref{FigMNv2} (Fig.~\ref{FigMNlam5}), we show the region in the $\lambda_5=0$ ($m_{12}^2=0, \lambda_5\ne 0$) $\nu$2HDM
where these scatterings are still in equilibrium at $T\lesssim M_1/3$.\footnote{A similar
plot to Fig.~\ref{FigMNv2} appears in Ref.~\cite{Haba2011ra} in the context of the Type I $\nu$2HDM with $v_2> 0$.
We are not aware of any plot similar to Fig.~\ref{FigMNlam5} in the literature, though
see Refs.~\cite{Ma2006fn,Hambye2009pw,Suematsu2011va,Kashiwase2012xd,Racker2013lua} 
for leptogenesis studies at points in the Ma model parameter space.}
This is the region where strong $\Delta L = 2$ scatterings can potentially wash out the generated asymmetry,
depending on the details of the leptogenesis 
(e.g. in a weak washout scenario with $N_1$ decays at $T\ll M_1$ this washout may be avoided).
Demanding that the scatterings fall out of equilibrium 
before sphaleron freeze-out at $T\sim 100$~GeV
provides a lower bound $v_2\gtrsim 0.3$ or $\lambda_5\gtrsim 10^{-5}$;
this is represented by the strong $\Delta L = 2$ scattering washout regions 
in Figs.~\ref{FigMNv2} and \ref{FigMNlam5} respectively.
We note that this calculation has been performed in the context of a perturbative theory.
This is reliable for the Type I $\nu$2HDM but not
for Type II, LS, or Flipped $\nu$2HDMs with sufficiently small $v_2$,
when perturbativity breaks down.

Putting this all together, we can now read off from 
Figs.~\ref{FigMNv2} and \ref{FigMNlam5} the regions of parameter space
which can achieve natural hierarchical thermal leptogenesis.
For $\nu$2HDMs with $m_{12}^2>0$ and $\lambda_5=0$, 
we find $10^3\text{ GeV}\lesssim M_{N_1}\lesssim \text{few} \times 10^7\text{ GeV}$ is viable for Type I $\nu$2HDMs, 
and $10^4\text{ GeV}\lesssim M_{N_1}\lesssim \text{few} \times 10^7\text{ GeV}$ 
for all other Types if they are to remain perturbative.
For the Ma model with $m_{12}^2=0$ and $\lambda_5\ne 0$, we find viable parameter space for
$10^3\text{ GeV}\lesssim M_{N_1}\lesssim 10^8\text{ GeV}$ and $10^{-5}\lesssim \lambda_5 \lesssim 0.5$

Lastly we note that the lightest scalar state in the Ma model is stable.\footnote{The
lifetimes of the heavier scalar states are governed by mass splittings $\Delta$ via
$\Gamma \sim G_F^2\Delta^5/(10^2 \pi^3)$.
In the parameter space of interest, one can check that $\Delta$ is
typically already large enough at tree-level so that lifetimes remain
well below $\mathcal{O}(1$ s) and therefore do not disturb big bang nucleosynthesis.}
It is therefore possible that this state, if it is neutral,
constitutes some or all of the observed dark matter.
During the leptogenesis epoch, $\Phi_2$ is produced in abundance in $N_1$ decays.
Overproduction of dark matter is of no concern 
as long as $\Phi_2$ efficiently thermalises at or below the temperatures when $N_1$ decays occur,
which suggests $m_{22}\ll M_{N_1}$.
In this case the lightest state is a thermal relic dark matter candidate.

\section{Conclusion \label{secconclusion}}

The minimal Type I see-saw model is unable to explain neutrino masses and
the BAU via hierarchical thermal leptogenesis without ceding naturalness.
The main conclusion of this paper is the observation that a second Higgs doublet can avoid this problem.
These $\nu$2HDM models provide a natural solution by reducing the (possibly effective) vev entering the see-saw formula.
This can be done radiatively, 
or by having the second Higgs doublet provide a tree-level see-saw with a small vev $v_2$,
kept natural by softly breaking a $U(1)$ or $Z_2$ symmetry.

The models naturally accommodate a SM-like Higgs
and predict the existence of approximately TeV-scale extra scalar states in order to remain natural.
We rediscovered the radiative Ma model as the only possibility when $v_2=0$;
in that case we found $10^3\text{ GeV}\lesssim M_{N_1}\lesssim 10^8\text{ GeV}$
and $10^{-5}\lesssim \lambda_5 \lesssim 0.5$
could simultaneously explain neutrino masses and the BAU via leptogenesis while remaining natural.
The $v_2 > 0$ models require $\tan\beta \gtrsim 4$;
we found $10^3\text{ GeV}\lesssim M_{N_1}\lesssim \text{few} \times 10^7\text{ GeV}$
was viable for Type I $\nu$2HDMs, and $10^4\text{ GeV}\lesssim M_{N_1}\lesssim \text{few} \times 10^7\text{ GeV}$
for all other Types if they are to remain perturbative.
The interesting areas of parameter space are well summarised in Figs.~\ref{FigMNv2} and \ref{FigMNlam5}.

\acknowledgments

This work was supported in part by the Australian Research Council.

\clearpage

\appendix
\begin{widetext}

\section{Squared scalar masses at $\mathcal{O}(m_{12}^4/m_{22}^4)$ \label{AppMasses}}

To order $m_{12}^4/m_{22}^4$, the scalar masses Eqs.~\ref{Eqscalarmasses} are given by
\begin{align}
 m_h^2 		&\approx v_1^2\left[ \lambda_1 + 
  \frac{m_{12}^4}{m_{22}^4} 
  \frac{2\lambda_{345}-\lambda_1-\frac{v_1^2}{2m_{22}^2}\lambda_1\lambda_{345}}
       {\left(1+\frac{v_1^2}{2m_{22}^2}\left(\lambda_{345}-2\lambda_1\right)\right)
        \left(1+\frac{v_1^2}{2m_{22}^2}\lambda_{345}\right)^2}
 \right], \nonumber \\
 m_H^2 		&\approx m_{22}^2\left[ 1 + \lambda_{345}\frac{v_1^2}{2m_{22}^2} +
 \frac{m_{12}^4}{m_{22}^4} 
  \frac{1-\frac{v_1^2}{2m_{22}^2}\left(2\lambda_{345}-3\lambda_2\right)
        +\frac{v_1^4}{4m_{22}^4}\left(\lambda_{345}^2+3\lambda_2\lambda_{345}-6\lambda_1\lambda_2\right)}
       {\left(1+\frac{v_1^2}{2m_{22}^2}\left(\lambda_{345}-2\lambda_1\right)\right)
        \left(1+\frac{v_1^2}{2m_{22}^2}\lambda_{345}\right)^2}
 \right], \nonumber \\
 m_A^2		&\approx m_{22}^2\left[ 1 + (\lambda_{345}-2\lambda_5)\frac{v_1^2}{2m_{22}^2}  + 
 \frac{m_{12}^4}{m_{22}^4}
  \frac{1+\frac{v_1^2}{2m_{22}^2}\left(\lambda_{345}+\lambda_2-2\lambda_5\right)}
       {\left(1+\frac{v_1^2}{2m_{22}^2}\lambda_{345}\right)^2}
 \right], \nonumber \\
 m_{H^\pm}^2	&\approx m_{22}^2\left[ 1 + \lambda_3\frac{v_1^2}{2m_{22}^2}  + 
 \frac{m_{12}^4}{m_{22}^4}
  \frac{1+\frac{v_1^2}{2m_{22}^2}\left(\lambda_2+\lambda_3\right)}
       {\left(1+\frac{v_1^2}{2m_{22}^2}\lambda_{345}\right)^2} 
 \right]. \label{EqscalarmassesApp}
\end{align}
\end{widetext}

\section{Bounds for more general $m_\nu$ \label{AppBounds}}

We consider the see-saw Lagrangian as in Eq.~\ref{EqSeesawYuk}
and a neutrino mass matrix of the form
\begin{align}
 m_{\nu} = \frac{v^2}{2} y_\nu \mathcal{D}_M^{-1} \mathcal{D}_{f(M)} y_\nu^T,
\end{align}
where $\mathcal{D}_x \equiv \text{diag}(x_1,x_2,x_3)$.
Note that in the $\nu$2HDM with $\lambda_5=0$ we have $f(M_j)=v_2^2/v^2$, 
and in the Ma model with $M_N\gg m_{22}$ we have
\begin{align}
 f(M_j) = \frac{\lambda_5}{8\pi^2}\left(\ln\left[\frac{M_j^2}{(m_H^2+m_A^2)/2}\right]-1\right).
\end{align}
Following Casas-Ibarra \cite{Casas2001sr}, it is possible to write
\begin{align}
 y_\nu = \frac{\sqrt{2}}{v}U^\dagger \mathcal{D}_m^\frac12 R \mathcal{D}_M^\frac12 \mathcal{D}_{f(M)}^{-\frac12},
\end{align}
where $R$ is a (possibly complex) orthogonal ($R R^T = R^T R = \mathbb{I}$) matrix.
The Vissani bound on each right-handed neutrino mass becomes \cite{Clarke2015gwa}
\begin{align}
 \frac{1}{4\pi^2}\frac{2}{v^2} \frac{M_j^3}{f(M_j)} \sum_i m_i |R_{ij}|^2 < 1\text{ TeV}^2 \nonumber \\
 \Rightarrow M_{N_1} \lesssim 3\times 10^7\text{ GeV} \times f(M_{N_1})^\frac13.
\end{align}
The CP asymmetry for hierarchical neutrinos \cite{Davidson2002qv} becomes
\begin{align}
 |\epsilon_1| & = \frac{6}{8\pi}\frac{M_1}{v^2}
 \frac{\text{Im}[(R^\dagger \mathcal{D}_m (R\mathcal{D}_{f(M)}^{-1}R^T)\mathcal{D}_m R^*)_{11}]}{(R^\dagger \mathcal{D}_m R)_{11}} \nonumber \\
  & \lesssim \frac{6}{8\pi}\frac{m_3 M_{N_1}}{v^2} \frac{1}{\text{min}[f(M_j)]} ,
\end{align}
where the approximate inequality holds for max$(|R_{ij}|)\le 1$.
For larger max$(|R_{ij}|)$ the inequality can be exceeded, 
but this corresponds to a fine-tuning (see Ref.~\cite{Clarke2015gwa}).
With this caveat, the Davidson-Ibarra bound for $\text{min}[f(M_i)]=f(M_{N_1})$ therefore becomes
\begin{align}
 M_{N_1} & \gtrsim  5\times 10^8\text{ GeV} \times f(M_{N_1}) .
\end{align}
The $\Delta L=2$ scatterings are proportional to \cite{Buchmuller2004nz,Buchmuller1997yu}
\begin{align}
 \sum_{i,j} & \text{Re}[(y_\nu^\dagger y_\nu)_{ij} (y_\nu^\dagger y_\nu)_{ij}] \frac{1}{M_i M_j} \nonumber \\
  & = \text{Tr}[ (y_\nu \mathcal{D}_M^{-1} y_\nu^T) (y_\nu \mathcal{D}_M^{-1} y_\nu^T)^\dagger ] \nonumber \\
  & = \frac{4}{v^4}
        \text{Tr}[ \mathcal{D}_m (R \mathcal{D}_{f(M)}^{-1} R^T) \mathcal{D}_m (R \mathcal{D}_{f(M)}^{-1} R^T)^\dagger ] \nonumber \\
  & \lesssim \frac{4}{v^4} \frac{\overline{m}^2}{\text{min}[f(M_j)]^2},
\end{align}
where the last line is an exact equality for $f(M_1)=f(M_2)=f(M_3)$,
an exact inequality when $R$ is real,
and an approximate inequality (as indicated) if $R$ is complex with max$(|R_{ij}|)\le 1$.
Again, for larger max$(|R_{ij}|)$ the inequality can be exceeded.
Then for $\text{min}[f(M_i)]=f(M_{N_1})$, the $\Delta L=2$ scattering Eq.~\ref{EqDL2Scat} becomes
\begin{align}
  \frac{\Gamma_{\Delta L=2}}{H} \lesssim \frac{T}{2.2\times 10^{13}\text{ GeV}}
  \frac{1}{f(M_{N_1})^2} \left(\frac{\overline{m}}{0.05\text{ eV}}\right)^2 .
\end{align}

\bibliography{references}

\end{document}